\documentclass[11pt]{llncs}

\usepackage{amssymb,amsmath,amsfonts}
\usepackage{verbatim}
\usepackage{graphicx}
\usepackage{xcolor}
\usepackage{dblfloatfix}
\usepackage{xspace}
\usepackage[printSOfinal=false]{pdfcomment}
\usepackage{multirow}
\makeatletter
\newif\if@restonecol

\usepackage[ruled,vlined,,commentsnumbered]{algorithm2e}
\usepackage{tikz}
\usepackage{hyperref}
\usetikzlibrary{arrows,decorations,shadows,positioning,shapes,backgrounds,fit,automata,calc,topaths,plotmarks}

\tikzset{mydesign/.style={
  rectangle, rounded corners,
  minimum size=11mm,
  minimum height=11mm,
  very thick,
  draw=#1!50!black!50, 
  top color=white, 
  bottom color=#1!50!black!20, 
  drop shadow={shadow xshift=.7ex, shadow yshift=-.7ex},
  text badly centered, anchor=north, text=black!70, text width=1.5cm
 },mydesign/.default=white,
 myrect/.style={
  rectangle,
  draw,
  minimum width=2cm,
  minimum height=1cm,
  thick,
  top color=white, 
  bottom color=#1!50!black!20, 
 },
 myrect/.default=white,
 interface/.style={
  rectangle,
  inner sep=.3ex,
  very thin,
  draw=#1!50!black!50,
  top color=white,
  bottom color=#1!50!black!20,
  node distance=0,
  font=\tiny
 },
 interface/.default=white,
  >=stealth',thick
}

\tikzset{mycircle/.style={
  mydesign=#1,
  circle
},mycircle/.default=blue}

\tikzset{myellipse/.style={
  mydesign=#1,
  ellipse
},myellipse/.default=blue}

\tikzset{mux/.style={
style=trapezium,trapezium stretches body,draw,thick,minimum width=2cm,minimum height=0.4cm,trapezium angle=35
}}

\tikzstyle{solution}=[mydesign=green]
\tikzstyle{threat}=[mydesign=orange]
\tikzstyle{action}=[mydesign=blue]

\tikzstyle{decision_node} = [diamond,
  very thick, 
  draw=#1!50!black!50, 
  top color=white, 
  bottom color=#1!50!black!20,
  drop shadow={shadow xshift=.7ex, shadow yshift=-.7ex},
  text width=4.5em, 
  text badly centered,
  node distance=3cm,
  inner sep=0pt]

\tikzstyle{decision} = [decision_node=blue]

\tikzset{
  mydoublearrow/.style={
    style=double arrow,
    color=black!70,
    draw,
    fill=#1!30,
    line width=.08cm,
    inner xsep=0.3cm,
    inner ysep=0.1cm},
  mydoublearrow/.default=white
}
\makeatletter

\pgfdeclareshape{reg}{
  \savedanchor\northeast{%
    \pgfmathsetlength\pgf@x{\pgfshapeminwidth}%
    \pgfmathsetlength\pgf@y{\pgfshapeminheight}%
    \pgf@x=0.5\pgf@x
    \pgf@y=0.5\pgf@y
  }
  \savedanchor\southwest{%
    \pgfmathsetlength\pgf@x{\pgfshapeminwidth}%
    \pgfmathsetlength\pgf@y{\pgfshapeminheight}%
    \pgf@x=-0.5\pgf@x
    \pgf@y=-0.5\pgf@y
  }
  \inheritanchorborder[from=rectangle]

  \anchor{center}{\pgfpointorigin}
  \anchor{north}{\northeast \pgf@x=0pt}
  \anchor{east}{\northeast \pgf@y=0pt}
  \anchor{south}{\southwest \pgf@x=0pt}
  \anchor{west}{\southwest \pgf@y=0pt}
  \anchor{north east}{\northeast}
  \anchor{north west}{\northeast \pgf@x=-\pgf@x}
  \anchor{south west}{\southwest}
  \anchor{south east}{\southwest \pgf@x=-\pgf@x}
  \anchor{text}{
    \pgfpointorigin
    \advance\pgf@x by -.5\wd\pgfnodeparttextbox%
    \advance\pgf@y by -.5\ht\pgfnodeparttextbox%
    \advance\pgf@y by +.5\dp\pgfnodeparttextbox%
  }

  \anchor{CLK}{
    \pgf@process{\northeast}%
    \pgf@x=-1\pgf@x%
    \pgf@y=-.6\pgf@y%
  }
  \anchor{CE}{
    \pgf@process{\northeast}%
    \pgf@x=-1\pgf@x%
    \pgf@y=-0.33333\pgf@y%
  }
  \anchor{R}{
    \pgf@process{\northeast}%
    \pgf@x=0pt%
  }
  \anchor{S}{
    \pgf@process{\northeast}%
    \pgf@x=0pt%
    \pgf@y=-\pgf@y%
  }
  \backgroundpath{
    \pgfpathrectanglecorners{\southwest}{\northeast}
    \pgf@anchor@reg@CLK
    \pgf@xa=\pgf@x \pgf@ya=\pgf@y
    \pgf@xb=\pgf@x \pgf@yb=\pgf@y
    \pgf@xc=\pgf@x \pgf@yc=\pgf@y
    \pgfmathsetlength\pgf@x{1.2ex} 
    \advance\pgf@ya by \pgf@x
    \advance\pgf@xb by \pgf@x
    \advance\pgf@yc by -\pgf@x
    \pgfpathmoveto{\pgfpoint{\pgf@xa}{\pgf@ya}}
    \pgfpathlineto{\pgfpoint{\pgf@xb}{\pgf@yb}}
    \pgfpathlineto{\pgfpoint{\pgf@xc}{\pgf@yc}}
    \pgfclosepath

  }
}
\makeatother
\tikzset{add font/.code={\expandafter\def\expandafter\tikz@textfont\expandafter{\tikz@textfont#1}}}
\tikzset{every reg node/.style={draw,minimum width=2cm,minimum 
height=1cm,thick,inner sep=1mm,outer sep=0pt,cap=round,add 
font=\sffamily}}

\tikzset{every rect node/.style={draw,minimum width=2cm,minimum 
height=1cm,thick,inner sep=1mm,outer sep=0pt,cap=round,add 
font=\sffamily}}

\colorlet{myorange}{red!30!yellow}
\defineavatar{TF}{color=myorange,author={Thomas Feller}}
\defineavatar{SA}{color=yellow,author={Sami Alsouri}}
\defineavatar{SK}{color=red!80!white,author={Stefan Katzenbeisser}}

\begin{document}
{
	\title{Hardware-based Security for Virtual Trusted Platform Modules}
	\author{Sami Alsouri \and Thomas Feller \and Sunil Malipatlolla \and Stefan Katzenbeisser}
	\institute{
		Technische Universit\"at Darmstadt\\
		Center for Advanced Security Research Darmstadt - CASED\\
		Mornewegstra{\ss}e 32, 64293 Darmstadt, Germany\\
	}
}

\maketitle
\begin{abstract}
Virtual Trusted Platform modules~(TPMs) were proposed as a software-based
alternative to the hardware-based TPMs to allow the use of
their cryptographic functionalities in scenarios where multiple TPMs are required
in a single platform, such as in virtualized environments. However, virtualizing
TPMs, especially virutalizing the Platform Configuration Registers~(PCRs),
strikes against one of the core principles of Trusted Computing, namely
the need for a hardware-based root of trust. 
In this paper we show how strength of hardware-based security can be gained
in virtual PCRs by binding them to their corresponding hardware PCRs.
We propose two approaches for such a binding. For this purpose,
the first variant uses binary hash trees, whereas the other variant
uses incremental hashing. In addition, we present an FPGA-based implementation
of both variants and evaluate their performance.
%
%
\end{abstract}




\section{Introduction}
\label{sec:intro}
The TPM chip provides secure storage for Platform Configuration
Registers (PCRs), which are supposed to store integrity measurements
in a trustworthy manner.
Trusted Computing requires for PCRs to be recorded in shielded locations
within the TPM. This provides a hardware-based implementation for PCRs
and makes them resistant against software attacks.

Unfortunately, the concept of one hardware TPM for every platform
is not adequate in scenarios where multiple TPMs are needed on
the same platform, such as virtualization scenarios.
To solve this problem, the concept of virtual TPMs~(vTPMs)~\cite{vtpm}
was proposed to allow the utilization of TPM functionalities,
such that each virtualized system is associated to an isolated TPM instance.
vTPMs are currently implemented in software.

However, virtualizing TPMs brings some important security challenges
and problems. First, virtualization causes the loss of hardware-based
security of TPMs. That is, virtual PCRs (vPCRs) are prone to software
attacks, which one tried to avoid using hardware TPMs.
Second, virtualizing TPMs increases the size of the Trusted Computing Base~(TCB).
Therefore, hardware-based security for virtual PCRs is preferable.

Current approaches for virtualizing TPMs~\cite{vtpm,vTPM2007,propertyvTPM} do not
provide -- to the best of our knowledge -- hardware-based security for vPCRs. 
To gain the strength of hardware-based security and to reduce the TCB,
we propose in this paper an approach to bind vPCRs to hardware PCRs.
More specifically, we provide two variants for this binding; the first uses binary
hash trees~\cite{hashtree} and the second uses the concept of incremental
hashing~\cite{incHashing97,Goi2001}. In the first variant all vPCRs of the
same index -- on a platform -- are jointly hashed using binary hash
trees. The root hash value is stored in the hardware PCR.
In the second variant, we use the incremental hashing approach,
so that an aggregated hash value can be stored in the hardware TPM chip.
%

Both approaches require the calculation of the hash tree or the incremental
hash inside the TPM to guarantee the security of the hash result.
Unfortunately, the current TPM specification does not provide interfaces for
such operations. Thus, we propose some additions to the TPM sepecifictions.
While it is difficult to change deployed TPM chips, next-generation
TPMs~\cite{tpm.next} will allow specifying required cryptographic
functionalities; furthermore, the concept of reconfigurable
TPM chips~\cite{EisenbarthTPM,GlasTPM,tinyTPM} allows the implementation of new functionalities
with relative ease.
%

We implement both approaches using a Virtex5 FPGA platform and
show that the application of both approaches
can increase the security of virtual TPMs with reasonable overhead.

This paper is structured as follows: Section~\ref{sec:back} gives a brief background
about Trusted Computing.
In Section~\ref{sec:approach}, we present our approaches to bind
virtual PCRs to hardware PCRs. Section~\ref{sec:impl} describes
our implementation of both approaches.
The evaluation of the approaches and their implementations is carried out
in Section~\ref{sec:eval}. Finally, we conclude in Section~\ref{sec:conc}.

\section{Background \& Related Work}
\label{sec:back}
The standards and specifications of Trusted Computing~(TC)~\cite{TPM2011},
a technology developed by the Trusted Computing Group~(TCG)\footnote{http://www.trustedcomputinggroup.org/},
provide many functionalities to secure computing platforms. 
TC relies on a cryptographic module,
called TPM~\cite{TPM2011}, that provides various security
functionalities. A TPM is a microcontroller-based chip with
hard-wired engines for various cryptographic functions such as RSA,
SHA-1 and HMAC. It forms the trust anchor of a system by building
a chain-of-trust which includes all loaded software on the platform.
The chain is extended based on the principle~\emph{hash then load}:
the executable code of every loaded software is hashed using the SHA-1 algorithm before
passing control to it. The computed hash values, representing
the state of the system, are stored in the PCRs of the TPM.



Remote Attestation is one important security function provided
by the TPM. In remote attestation, the system equipped with a TPM
trustworthily reports its platform state to a remote challenger.
For this, the TPM provides a set of PCR values signed by an
Attestation Identity Key (AIK) and a Stored Measurement Log~(SML)
to the challenger. In turn, the challenger decides on
the trustworthiness of the system by comparing them with
well-known reference values stored in a public Reference
Measurement List~(RML).


A conventional TPM is, in general, an Application Specific Integrated
Circuit~(ASIC)~\cite{TPM2011} implementation and therefore cannot
be updated after deployment.
However, there exist approaches in literature for supporting a flexible
update of cryptographic algorithms on the TPM using the reconfiguration
technology such as Field Programmable Gate Array (FPGA) as proposed by
Malipatlolla et al. in~\cite{stpm}.


Unfortunately, the current specification of the TPM does
not support hardware-based security for systems using
virtualization and cloud computing technologies.
Though there exist in literature designs supporting
resource constrained embedded systems~\cite{tinyTPM}
and arbitrary number of virtual TPMs~\cite{vTPM2007},
they do not address the above problem. Virtual TPMs
in these approaches belong therefore to the TCB of
a platform. 

The concept of hash trees has been used in many different contexts.
In the area of Trusted Computing, hash trees were applied
in~\cite{optimalPara} to protect memory regions using the region block size
and the number of memory updates as parameters for the hash tree.
Schmidt et al.~\cite{treeFormed} used hash trees during the
integrity measurement process to create tree-formed measurements, in which the measured
components represent the leaves and the PCR values
represent the roots. The goal of this work was
to allow detecting the position of a possible
manipulation of an SML, which was possible in case of using
linear ordered measurements~(like in TCG standard) only by checking
the integrity value of each entry in the SML.
Another work applied the concept of hash trees in TC
is the one presented by Sarmenta et al.~\cite{virtualCounters}.
The objective of the authors was to create
very large number of virtual monotonic counters on
an untrusted machine with a TPM. The virtual counters
can be then used to detect illegitimate modifications to shared
data objects~(including replay attacks and forking attacks)~\cite{offlineStorage}.
The authors proposed for this the use of additional
TPM commands to in order to calculate hash tree node and
root values in a secure manner.
However, we apply hash trees in our approach to bind virtual PCRs
to hardware PCRs, which is a security problem of virtual TPMs and therefore
we fulfill other purposes.
%
%
%

\section{Approach}
\label{sec:approach}
To provide hardware-based security for virtual PCRs, we propose
in this section two different approaches. The first uses the well-known
binary hash trees and the second uses incremental hashing. Both
approaches are based on the idea of binding all virtual PCRs with a specific
index to the hardware PCR of the same index in such a way that any
manipulation of a virtual PCR can be detected by the help of the value
of its corresponding hardware PCR.
\subsection{Hash Tree Based Binding}
\begin{figure}
\centering
\resizebox{0.6\columnwidth}{!}{\begin{tikzpicture}[level/.style={sibling distance=50mm/#1}]
\begin{scope}[grow=up]
\node [circle,draw] (z){$h_0$}
  child {node [circle,draw] (a) {$h_2$}
    child {node [circle,draw] (b) {$h_6$}
      child {node {$\vdots$}
        child {node [rectangle,draw,xshift=3mm] (d) {$vPCR_i^n$}}
        child {node [rectangle,draw,xshift=-3mm] (e) {$vPCR_i^{n-1}$}}
      } 
      child {node {$\vdots$}}
    }
    child {node [circle,draw] (g) {$h_5$}
      child {node {$\vdots$}}
      child {node {$\vdots$}}
    }
  }
  child {node [circle,draw] (j) {$h_1$}
    child {node [circle,draw] (k) {$h_4$}
      child {node {$\vdots$}}
      child {node {$\vdots$}}
    }
    child {node [circle,draw] (l) {$h_3$}
      child {node {$\vdots$}}
      child {node (c){$\vdots$}
        child {node [rectangle,draw,xshift=2mm] (o) {$vPCR_i^2$}}
        child {node [rectangle,draw,xshift=-2mm] (p) {$vPCR_i^1$}}
      }
    }
  };
\end{scope}

\path (a) -- (j);
\path (b) -- (g);
\path (k) -- (l);
\path (k) -- (g);
\path (d) -- (e);
\path (o) -- (p);
\path (o) -- (e) node (x) [midway] {$\cdots$};
\end{tikzpicture}
}
\caption{Sample Binary Hash Tree}
\label{fig:hash_tree}
\end{figure}
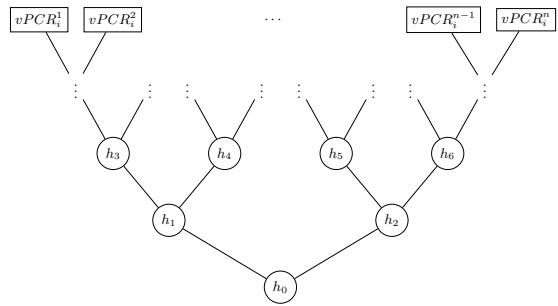

We propose the use of the concept of binary hash trees,
as shown in Figure~\ref{fig:hash_tree}. In the following, we explain our
approach using three phases; the setup phase, the integrity
measurement phase, and finally the remote attestation phase.

\paragraph{Setup Phase.}
We construct the hash tree in the following way: The leaves
at the top of the tree present all vPCRs of a specific index~$i$
of all existing vTPMs ($1,\ldots,n$) on a platform. For instance, $vPCR_{10}^1$
indicates the vPCR number 10 of the vTPM number 1.
To increase efficiency, we propose using hash trees of fixed height $l$.
That is, with $l=10$, one can run 1024 vTPMs on the same platform 
bound to a single hardware TPM.
This number is probably enough for single platforms~(e.g., servers),
in case of using isolated vTPMs for virtual machines.
Nodes further down in the tree are the hashes of their respective
child nodes. Figure~\ref{fig:hash_tree} illustrates this process;
$h_0$ represents the accumulated vPCR values (root hash node)
that will be stored in the hardware TPM;
$h_0$ is obtained by combining the hashes $h_1$ and $h_2$, i.e., 
\begin{equation*}
h_0 \Leftarrow  hash(h_1 || h_2), 
\end{equation*}
where $||$ indicates the concatenation operation.
Similar to $h_0$ all intermediate hashes are computed.
That is, the calculation of $h_0$ depends on the calculation
of the leaves and all intermediate nodes in the hash tree.
Consequentially, any manipulation to one of the leaves
can be detected.
\paragraph{Integrity Measurement.}
%
Once a vPCR value needs to be updated, the vTPM is notified about the new
measurement and the new value of the vTPM is bound
to the corresponding hardware PCR as explained in the setup
phase. In addition, the SML of this vTPM is also updated.
%
%

More specific, the TSS notifies the underlying hardware TPM
by starting the procedure depicted in Algorithm~\ref{alg:hashStart},
sending the old vPCR value $vPCR_{old}$, the new vPCR value $vPCR_{new}$,
the height of the hash tree $l$ and the PCR index $i$ of the hardware
TPM that needs to be updated. This algorithm is then executed inside
the hardware TPM, which in turn stores these provided values in
temporary registers in the volatile storage. The algorithm returns
$OK$ if and only if the process was successfully finished and there
is no hash tree updating process currently running for this $PCR_i$.
\begin{algorithm}[t]
\KwIn{old vPCR value $vPCR_{old}$,\newline new vPCR value $vPCR_{new}$,\newline hardware PCR index $i$,\newline height of the tree $l$}
\KwOut{$OK$ or $error$}
\If(\tcp*[f]{a hash tree execution is running}){$c_i \neq 0$}{
	\Return error\;
}
\Else{
  $c_i \gets l$ \tcp*[f]{initialize counter with tree height}\newline
  $tmp_{old} \gets vPCR_{old}$\;
  $tmp_{new} \gets vPCR_{new}$\;
  \Return OK\;
}
\caption{\label{alg:hashStart}TPM\_Update\_Leaf\_Init}
\end{algorithm}


\begin{algorithm}[t]
\KwIn{hardware PCR index $i$, sibling}
\KwOut{updated hardware PCR value $PCR'_{i}$ or $error$}
$tmp_{old} = hash(tmp_{old}||sibling)$\;
$tmp_{new} = hash(tmp_{new}||sibling)$\;
$c_i \gets c_i-1$\;
\If(\tcp*[f]{root of tree reached}){$c_i = 0$}{
\If{$tmp_{old} = PCR_i$}{
$PCR_i \gets tmp_{new}$\;
\Return $PCR_{i}$\;
}
\Else{
\Return $error$; \tcp*[f]{the hash tree is tampered}\
}}

\caption{\label{alg:hashUpdate}TPM\_Update\_Leaf}
\end{algorithm}
After providing the hardware TPM with the old and new vPCR values,
re-calculation of the hash tree is required as shown
in Algorithm~\ref{alg:hashUpdate}.
Since the hash tree is located outside of the TPM and the algorithm
must be provided with all siblings located in the way to the root of
the hash tree, Algorithm~\ref{alg:hashUpdate} must be called
$l-1$ times, providing at each time the correct sibling of the
current hash tree level. First, the TPM is provided with the
sibling of the leaf (i.e., the vPCR) and hashes the old
and the new value of the vPCR with its sibling. The same process
is repeated until the root is reached (i.e., the tree height equals 0).
If the old vPCR value equals the value stored in $PCR_i$,
the hash tree is untampered and the newly calculated root
can be stored in $PCR_i$, otherwise an $error$ is returned, indicating
a potential software attack aiming at manipulating the PCR
values.
It does not matter who calls Algorithm~\ref{alg:hashUpdate}
in the hardware TPM to provide the siblings values, more important
is the provision of the correct values to calculate a correct
root value, which equals $PCR_i$. That is, an attacker which
calls Algorithm~\ref{alg:hashUpdate} after the Algorithm~\ref{alg:hashStart}
was called, would have to deliver a collision to $PCR_i$ in order to
successfully perform an attack on the TPM in order to update
the root value to another selected one. This is assumed to be hard
when using a collision-resistant hash function.

Note that it would be possible to provide the TPM with all required
siblings (from a leaf to the root) at once. Although this would reduce
the communication overhead with the TPM, it would
require at the same time the presence of enough temporary storage
for all these values, which could be a problem for resource
constraint TPM implementations.
\paragraph{Remote Attestation.}
The remote attestation process is very similar to the one described
by TCG, with one more difference, which is verifying the hash tree.
In details, after sending a nonce and signing it together with the requested
vPCR by a vAIK of a particular vTPM, the nonce is then forwarded to
the hardware TPM. The hardware TPM also signs the nonce and the value
of the requested PCR (i.e., the root node of the hash tree). The
signatures, the SML and the hash tree are finally sent to the challenger.
The challenger verifies the signatures and the SML.
In addition, the challenger re-calculates the hash tree as described
above. If the signed root value equals the re-calculated value~(which
means that the vPCR is untampered), the signed value can be considered
trusted.

\subsection{Incremental Hash Based Binding}
Incremental hashing is another efficient way to
aggregate hash values of messages that change over time.
More specific, an incremental hash function produces
an updated hash value of a modified message faster
than recomputing the hash from scratch.
We propose here an approach which uses incremental hashing
to aggregate all vPCR values of a platform and 
update the aggregated value after every extend operation
performed on any vTPM on the platform.

Algorithm~\ref{alg:incHash} details the hash update
procedure based on the incremental hashing scheme of~\cite{incHashing97}. 
Modular multiplication was chosen as the combining operation.
To comply to TCG standards, the updated hash will include
the history~($PCR_i$) of all measurements.
\begin{algorithm}
\KwIn{hardware PCR index $i$, old hash-value $vPCR_{old}$, new hash-value $vPCR_{new}$}
\KwOut{updated hardware PCR value $PCR'_i$}
$h_i = mod\_div(PCR_i, hash(i||vPCR_{old}))$\;
$PCR'_i = mod\_mult(h_i, hash(i||vPCR_{new}||PCR_i))$\;
\Return $PCR'_i$\;
\caption{\label{alg:incHash}TPM\_Increment\_Hash}
\end{algorithm}
%
\paragraph{Setup Phase.}
The incremental hash-based binding approach presented
herein can be used with an arbitrary number of vTPMs.
Adding and removing PCR values of vTPMs is done by multiplying/dividing
the corresponding hash values with the aggregated hardware 
PCR value.
To bind all vPCRs of a vTPM to the value of $PCR_i$, all
corresponding $vPCR_i$ of each vTPM are combined according
to the following equation, where $m$ is the prime modulus,
$i$ is the number of the PCR register and $n$
is the number of vTPM existing on the same platform:
\begin{equation*}
	PCR_i \Leftarrow \prod \limits_{k=1}^{n} hash(k||vPCR_i^k) \ mod\ m 
\end{equation*}
\paragraph{Integrity Measurement.}
For continuous integrity measurement the update of a
PCR value is performed according to Algorithm~\autoref{alg:incHash}.
In addition, $PCR_i$ is included in the updated value
$PCR'_i$. This is very important to do in order to avoid resetting a
PCR value and to keep track of the update history of a PCR.
\paragraph{Remote Attestation.}
The verification of the remote attestation process has
to include the integrity verification of the incremental hash.
In addition to the SML provided by a TSS of a vTPM,
an SML for the incremental hash updates is provided.
As defined by TCG, a challenger first verifies all
signatures and the SML of the vTPM. In addition,
the challenger uses the SML of the hardware TPM,
which has all incremental hash updates, to verify
the integrity of vTPM itself.

\section{Implementation}
\label{sec:impl}
\subsection{Hash Tree Based Binding}
\label{sec:ht}
To evaluate the feasibility of our approach we first implemented
the hash-tree based measurement scheme in hardware.
For fair comparison we throttled our design to comply to the TPM
specifications, although our FPGA-based implementation is able to
operate at a higher frequency.
An off-the-shelf TPM is running at 33MHz whereas our SHA-1
implementation is operating at a maximum frequency of 128.7MHz
on a Xilinx Virtex5 FPGA.
We assume that the implementation of the SHA-1 algorithm present
in the TPM features a similar performance as our straight
forward implementation.

Another important factor that is limiting the performance is that currently the TPMs are connected to the system via the LPC-Bus (Low Pin Count) \cite{LPC_intel}.
Therefore, we calculated the ideal rates for LPC-Bus transfers of the hash values from the software stack to a TPM.
The width of the LPC-Bus is 4 bit which leads to the fact, that the transfer of a 20 byte hash value takes 40 clock cycles without overhead.
According to the utilized transfer mode the overhead and total number of clock cyles is depicted in \autoref{tab:clockcycles}.
Although the DMA transfer is often not implemented for current TPMs we included the slower I/O write for completeness.
Note that the number of clock cycles stated in \autoref{tab:clockcycles} denotes only the time required for the transmission of one hash value. 
 
\begin{table}[b]
\caption{Clock Cycles for Hash Value Transmission}
\label{tab:clockcycles}
\centering
\begin{tabular}{l|c|c|c|l}
\hline
Mode 	& Hash 	& Overhead		& Total & Time \\
	&	& (cf.\cite{LPC_intel}) &	& @33MHz\\
\hline
I/O Write & 40 & 20*11=220 & 260 & 7.88$\mu$s\\
DMA Write & 40 & 5*24=120 & 160 & 4.84$\mu$s\\
\hline
\end{tabular}
\end{table}

\begin{figure}[h]
\centering
\resizebox{0.7\columnwidth}{!}{\includegraphics{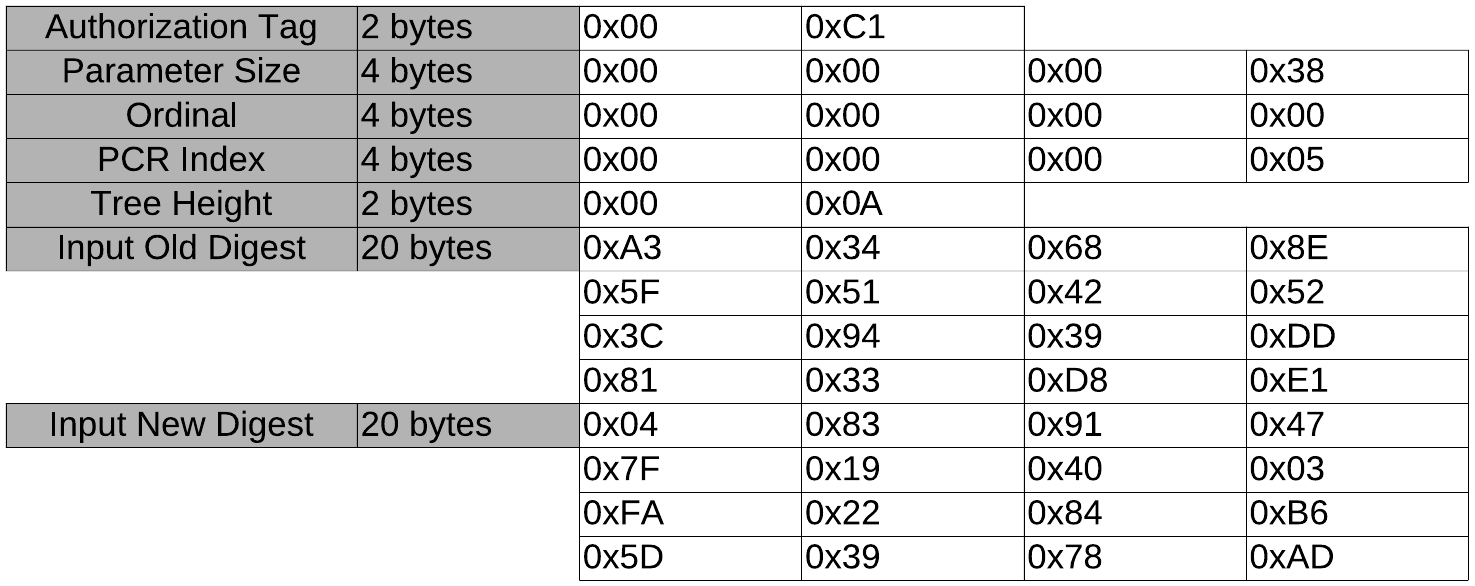}}
\caption{TPM\_Update\_Leaf input message block}
\label{fig:updateLeafInput}
\end{figure}

An overview of the \texttt{TPM\_Update\_Leaf\_Init} command structure is depicted in \autoref{fig:updateLeafInput}.
Each command is sent in 4 byte blocks with an overhead of 24 clock cycles.
The \texttt{TPM\_Update\_Leaf\_Init} message block size is 56 bytes, which results in a total transfer time of 112 + 14 * 24 = 448 clock cycles or 13.58$\mu$s.
A \texttt{TPM\_Update\_Leaf} message block consists of 34 bytes of data, which results in 8.61$\mu$s (68 + 9 * 24 = 284 clock cycles) transmission time.

We propose in this paper a specialized hardware architecture to
speed up the execution time of hash-tree computation by utilizing
two parallel SHA-1 modules.  Although it is possible to execute
the hash function in a serial manner, the actual speed of the
computation is of major importance in performance-critical
environments.

\begin{figure}[b!]
\centering
\resizebox{0.5\columnwidth}{!}{\begin{tikzpicture}

\node [shape=reg] (pcrold) {$PCR_{old}$};
\node [shape=reg,right=of pcrold] (sibling) {$sibling_i$};
\node [shape=reg,right=of sibling] (pcrnew) {$PCR_{new}$};

\node [mux,above=of pcrold,rotate=180] (oldinput) {};
\node [mux,above=of pcrnew,rotate=180] (newinput) {};

\node [above=of oldinput.205,yshift=-.5cm] (old){$PCR_{old}$};
\node [above=of newinput.335,yshift=-.5cm] (new){$PCR_{new}$};
\node [above=of sibling,yshift=.5cm] (sibl) {$sibling_i$};

\node [myrect,below=of pcrold.335,xshift=.5cm] (hashold) {SHA-1};
\node [myrect,below=of pcrnew.205,xshift=-.5cm] (hashnew) {SHA-1};

\node [below=of hashold] (oldprime) {$PCR'_{old}$};
\node [below=of hashnew] (newprime) {$PCR'_{new}$};

\draw [->] (old) -- (oldinput.205);
\draw [->] (new) -- (newinput.335);
\draw [->] (sibl) -- (sibling);

\draw [->] (hashold.south) |- ++(-3,-.4) |- ($(oldinput.335) + (0,.4)$) -| (oldinput.335);
\draw [->] (hashnew.south) |- ++(3,-.4) |- ($(newinput.205) + (0,.4)$) -| (newinput.205);

\draw [->] (oldinput) -- (pcrold);
\draw [->] (newinput) -- (pcrnew);

\draw [->] (pcrold) |- ($(hashold.135) + (0,.4)$) -- (hashold.135);
\draw [->] (pcrnew) |- ($(hashnew.45) + (0,.4)$) -- (hashnew.45);

\draw [->] (sibling.240) |- ($(hashold.45) + (0,.5)$) -| (hashold.45);
\draw [->] (sibling.300) |- ($(hashnew.135) + (0,.5)$) -| (hashnew.135);

\draw [->] ($(hashold.south) + (0,-.4)$) -- (oldprime);
\draw [->] ($(hashnew.south) + (0,-.4)$) -- (newprime);

\end{tikzpicture}}
\caption{Datapath for Hash-Tree}
\label{fig:ht-datapath}
\end{figure}
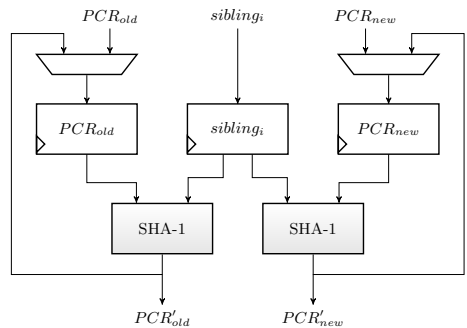

In \autoref{fig:ht-datapath} the parallel datapath of
our hash tree implementation is depicted. As soon as the
next sibling is written to the $sibling_i$ register and
the PCR registers contain the most recent values, the hash
generation of the supplied values begins.
In Algorithm~\ref{alg:hashUpdate} the sibling is
always appended to the temporary PCR contents and then
provided as an input to the hash function.

\subsection{Incremental Hash Based Binding}



For the incremental hashing approach the output of a
SHA-1 is too short.  Therfore, we utilize the SHA-2
512 bit variant to realize the incremental hashing function.
The modular multiplication is performed using an interleaving
multiplication algorithm as presented in \cite{CryptAlg}.
We refused to use the faster montgomery multiplication algorithm,
as the overhead for encoding/decoding the operands/results is
only bearable if a limited set of operands is used, such as
exponentiation. Therefore, the implementation of the shift
and add multiplication from \cite{deschamps2009} is used.

The computation of the hash value utilizes the multiplication algorithm, whereas the slower division algorithm is used only for updating.
The computational complexity of updating hash values is constant for the incremental hashing scheme, which counterbalances the operational deficit of binary division.
However the implementation of the algorithms presented in this paper can still be optimized for better performance.

The implementation of the incremental hashing approach cannot be transferred directly to an off-the-shelf TPM as done with the hash tree based approach (c.f. Sec.~\ref{sec:ht}), because hardware-based SHA-2, modular multiplication, and modular division implementations are missing.
Therefore, a comparison between an off-the-shelf TPM, supporting the incremental hashing approach and the hardware based implementation presented in this paper is not possible.

\section{Evaluation}
\label{sec:eval}
\paragraph{Binding using hash trees.}
The SHA-1 algorithm presented in this paper is a rather straight forward
design which was not optimized for performance as there are various SHA-1
FPGA implementations available in literature (cf.~\cite{CryptAlg}).
The overall computational time of an hash-tree update is depicted in~\autoref{tab:hashComp}.
To compare our implementation with the modification of a standard TPM,
which executes the SHA-1 computations in serial mode, we also listed
the expected results in~\autoref{tab:hashComp}.

\begin{table}[h]
\centering
\caption{Computational time for Hash Tree updates}
\label{tab:hashComp}
\begin{tabular}{|l|c|c|c|c|c|c|c|}
\hline
& & \multicolumn{3}{|c|}{Clock Cycles} & \multicolumn{2}{|c|}{Time}\\\cline{3-7}
Design & $h$ & SHA-1 & Command & Total & @33MHz & @128.7MHz \\
&&&Transmission&&&\\
\hline\hline
parallel & 2 & 2*175 & 448+2*284 & 1366 & 41.4$\mu$s&10.6$\mu$s\\
parallel & 10 & 10*175 & 448+10*284 & 5038 & 152.7$\mu$s&39.2$\mu$s\\
parallel & 20 & 20*175 & 448+20*284 & 9628 & 291.8$\mu$s&75.8$\mu$s\\
\hline\hline
serial & 2 & 4*175 & 448+2*284 & 1716 & 52$\mu$s&13.3$\mu$s\\
serial & 10 & 20*175 & 448+10*284 & 6788 & 205.7$\mu$s&52.7$\mu$s\\
serial & 20 & 40*175 & 448+20*284 & 13128 & 397.8$\mu$s&102.0$\mu$s\\
\hline
\end{tabular}
\end{table}

The parallel execution of the hash functions reduces the
computational overhead to the minimum.
The table clearly shows the bottleneck is the communication
over the LPC-Bus, which takes the most significant amount of time.
Choosing a faster SHA-1 implementation will only reduce
approximately 20\% of the number of clock cycles, as the
communication over the LPC-bus uses about 80\% of the clock cycles.
Therefore, in addition to the parallel implementation of
the hash tree scheme, the communication interface has to
be improved.
\paragraph{Binding using incremental hashing.}
In the following we give an estimation on the implementation
of an incremental hash based binding scheme.
\autoref{tab:incHash} summarizes the resource consumption
of the utilized algorithms.

\begin{table}
\centering
\caption{Resource consumption of Incremental Hashing}
\label{tab:incHash}
\begin{tabular}{l|r|r|r|r}
\hline
Scheme & LUTs & Registers & Frequency & Cycles \\
\hline
Shift and Add Mult. & 14371 & 6175 & 323.415MHz & 2053 \\
Binary Div. Algorithm & 4128 & 8430 & 32.275MHz & 1563 \\
SHA-512 & 1423 & 2744 & 128.617MHz & 81 \\
\hline
\end{tabular}
\end{table}

\paragraph{Comparison of both variants.}
A comparison of the overhead for the proposed schemes
is given in~\autoref{tab:overhead}, where $n$ represents
the number of vTPMs and $u$ represents the average number
of extend operations for each vTPM.
The table includes the order of complexity and the time
consumed per operation.
Although the update operation of the incremental hash-based
binding takes less time if there are more than 16 vTPMs in
use~(i.e., the hash tree height equals 4), the time consumed by
the verify operation is at least an order of magnitude
slower than the hash tree-based approach.

The main issue of the incremental hashing approach is to
keep track of the stored values in the hash.
The property of deleting values from the hash by division
is contrary to the trusted computing requirement, as one
could easily reset the stored hash.
To mitigate this property of incremental hashing, we included
the history of the content, as it is done in current TPMs.
This leads to the fact that the incremental hashing approach
still features a constant complexity for the mostly utilized
update function, but the verification complexity becomes
dependent on the number of updates.
\begin{table}
\centering
\caption{Overhead of different Schemes}
\label{tab:overhead}
\begin{tabular}{|c|c|c|c|}
\hline
& Hash-Tree & Incremental Hashing & \\
\hline\hline
read/verify & $O(log(n))$ & $O(n * u)$ & \multirow{2}{*}{complexity}\\
write/update & $O(log(n))$ & $O(1)$ & \\
\hline
read/verify & 1.4$\mu$s & 55.4$\mu$s & \multirow{2}{*}{time}\\
write/update & 1.4$\mu$s & 6.9$\mu$s & \\
\hline
\end{tabular}
\end{table}


\section{Conclusion}
\label{sec:conc}
The major contribution of this paper was to provide hardware-based
security to the virtual TPMs by binding them to a single hardware TPM.
For this, two novel approaches, hash tree based binding and
incremental hash based binding, have been proposed. 
Both variants have been implemented and evaluated on state-of-the-art
Virtex5 FPGA platform. Our evaluation shows that the update
process of a hash tree based binding is done with a complexity of
$O(log(n))$. The same process for an incremental hashing 
based binding has complexity $O(1)$. However, the verification
process of the incremental hashing based binding is much more
expensive than the hash tree based one. In general, our evaluation
shows that the approach is applicable with reasonable overhead.


%
\bibliographystyle{abbrv}
\bibliography{literature}  
%
%
\end{document}